\documentclass[11pt]{article}
\usepackage{epsf} 
\usepackage{amsmath}
\usepackage{amssymb}
\usepackage{epsfig}
\usepackage{latexsym}
\usepackage{amsfonts}
\usepackage{graphicx}%
\usepackage{varioref}
\usepackage{ifthen}
\begin{document}
\parindent 0mm 
\setlength{\parskip}{\baselineskip} 
\thispagestyle{empty}
\pagenumbering{arabic} 
\setcounter{page}{0}
\mbox{ }
\rightline{UCT-TP-298/13, MITP/13-056, IFIC/13-70}
\newline
\rightline{April 2014}
\newline
\vspace{0.1cm}
\begin{center}
{\Large {\bf 	$B$ Meson Decay Constants $f_{B_c}$, $f_{B_s}$ and $f_B$ from QCD Sum Rules}}
\end{center}
\vspace{.05cm}
\begin{center}
{\bf M. J. Baker} $^{(a)}$, {\bf J. Bordes} $^{(a)}$,
{\bf C. A. Dominguez} $^{(b)}$,\\
{\bf J. Pe\~{n}arrocha} $^{(a)}$,
 {\bf K. Schilcher}$^{(b),(c)}$
 \end{center}

\begin{center}
{\it $^{(a)}$ Departament Fisica Teorica and IFIC, Centro Mixto CSIC, Universitat de Valencia,  Calle Dr. Moliner 50, E-46100 Burjassot (Valencia), Spain}\\

{\it $^{(b)}$Centre for Theoretical Physics and Astrophysics and Department of Physics, University of
Cape Town, Rondebosch 7700, South Africa}\\

{\it $^{(c)}$ Institut f\"{u}r Physik, Johannes Gutenberg-Universit\"{a}t,
Staudingerweg 7, D-55099 Mainz, Germany}
\end{center}

\begin {footnotesize}
{\it E-mail}:baker.michael.james@googlemail.com, bordes@uv.es, cesareo.dominguez@uct.ac.za,
Jose.A.Penarrocha@uv.es, karl.schilcher@uni-mainz.de
\end{footnotesize}

\date{\today}

\begin{center}

\textbf{Abstract}
\end{center}
\noindent
Finite energy QCD sum rules with Legendre polynomial integration kernels are used to determine the heavy meson decay constant $f_{B_c}$, and revisit $f_B$ and $f_{B_s}$. Results exhibit excellent stability in a wide range of values of the integration radius in the complex squared energy plane, and of the order of the Legendre polynomial. Results are $f_{B_c} = 528 \pm 19$ MeV, $f_B = 186 \pm 14$ MeV, and $f_{B_s} = 222 \pm 12$ MeV. \\

\newpage
\section{Introduction}
The decay constant of a pseudoscalar meson $B_{q}$ consisting of a heavy
$b$-quark and a lighter
 $q$-quark, with $q=u,d,s,c$, is defined through the matrix
element of the pseudoscalar current
\begin{equation}
\langle\Omega|(m_{b}+m_{q})(\overline{q}i\gamma_{5}b)(0)|B_{q}
\rangle = f_{B_{q}}M_{B_{q}}^{2}\;,
\end{equation}
where $|\Omega\rangle$ is the physical vacuum, $M_{B_{q}}$ is the mass of the $B_{q}$ meson and $f_{B_{q}}$ the
corresponding leptonic decay constant.  These decay constants are of great phenomenological interest since they enter as input in non-leptonic $B$ decays, in the hadronic matrix elements of
$B-\bar{B}$ mixing, and in the extraction of CKM matrix elements $|V_{cb}|$,
$|V_{ub}|$ from the leptonic decay widths of $B$ mesons. The so-called hadronic $B $ parameter, which
is directly related to the deviation from the vacuum saturation hypothesis,  also requires knowledge of these leptonic decay constants, which  are of central interest to the ongoing experiments carried out at $B$ factories. Since their determination involves non-perturbative QCD effects, one has to rely
 on essentially two approaches, QCD sum rules (QCDSR) and lattice QCD (LQCD) simulations.
Since the pioneering work of Shifman, Vainshtein and Zakharov \cite{SHIFMAN} the sum rule method has been used successfully to calculate various
low energy parameters in QCD \cite{REV}. Specific sum rules are based on Borel transforms, Hilbert
transforms, positive moments or inverse moments. Sum rule calculations of the decay constants have been performed since the 1980's, with results in the range $f_{B}=160-230$ MeV, $f_{B_{s}}/f_{B}=1.1-1.4$, and $f_{B_c}=160 - 360$ MeV \cite{DOMINGUEZPAVER}-\cite{Colangelo:1992cx}.  A heavy quark effective theory
calculation \cite{PS} gives $f_{B}=206\pm20$ MeV, while historical LQCD
determinations fall in the wide range  $f_{B}=161-218$ MeV, and $f_{B_{s}}/f_{B}=1.11-1.16$ \cite{APE}. Recent QCDSR calculations have narrowed the range of results to
$f_{B}=189-207$ MeV, and $f_{B_{s}}=216-242$ MeV \cite{WANG13}-\cite{GEL13}, while LQCD calculations, claiming  high accuracy, give $f_{B}=186-205$
MeV, and $f_{B_{s}}=224-244$ MeV \cite{BAZ12}. For $B_{c}$ a compilation of many LQCD determinations \cite{Jones} gives values in the wide range $f_{B_{c}}=380- 680$ MeV. 
The experimental situation remains unclear. For instance, using the recent Belle result for $f_B$ \cite{Adachi:2012mm},
\begin{equation}
f_{B}|V_{ub}|=(7.4\pm0.8(\textrm{stat})\pm0.5(\textrm{syst}))\times10^{-4}\textrm{ GeV}\;,
\end{equation}
and $|V_{ub}|=(4.01\pm0.56) \times 10^{-3}$ \cite{Rosner:2013ica}, one obtains
\begin{equation}
f_{B}= 185\pm 35 \textrm{ MeV}
\end{equation}
However, a recent value of the branching fraction of $B^+ \rightarrow \tau^+ \; \nu$ from BaBar \cite{Lees:2012ju} gives $f_B = 295\; {\mbox{MeV}}$ ($f_B= 221\; {\mbox{MeV}}$), depending on their value of $V_{ub}$ from exclusive (inclusive) charmless semileptonic B-decays. \\
In a previous calculation of $f_{B}$ and $f_{B_{s}}$ by some of the present authors \cite{Bordes:2004vu}, a method was used based on finite energy QCDSR (FESR) which
equates positive moments of data with QCD theory. On the
theoretical side, a large momentum expansion in powers of $m_{b}^{2}/s$ was
taken from reference \cite{Chetyrkin:2001je}, where $m_{b}$ is the
mass of the bottom quark and $s$ the square of the center-of-mass energy. The perturbative
expansion was considered up to second order in the strong coupling constant,
and up to seventh order in $m_b^2/s$.  On
the phenomenological side, a combination of positive
moments involving Legendre polynomials was used to
optimize the experimental data, enhancing the lowest lying
$B_{q}$ meson. The contribution of the unknown continuum data
was shown to be  negligible, after a suitable choice of Legendre
polynomials \cite{Bordes:2004vu}. \\
The present paper is devoted to determining $f_{B_c}$, and updating results for $f_B$ and $f_{B_s}$. We use more recent information on the QCD side of the sum rule, and employ a new criterion for optimizing the stability of the result. For the QCD correlator we use an $O((m_b^2/s)^7)$ expansion of the pseudoscalar two-point function up to $O(\alpha_s^2)$ \cite{Chetyrkin:2001je}.  For $f_{B_c}$ and $f_{B_s}$ we supplement this with $O(m_q^4)$ light quark mass corrections up to $O(\alpha_s)$ \cite{Jamin:2001fw}.  To account for non-perturbative corrections we include terms up to dimension six in the operator product expansion (OPE).  Rather importantly, the correlator is expressed in terms of the running quark mass, rather than the pole mass \cite{Gray:1990yh}, as it is well known that this improves the convergence of the perturbative series \cite{Jamin:2001fw}.  
\section{Preliminaries}
To study the decay constants of $B_q$ mesons, with $q\in\{u, d, s, c\}$, we consider the pseudoscalar two-point correlator
\begin{equation}
\Pi(q^2)=i\int dx \, e^{iqx} \langle \Omega | T(j_5(x)j_5(0)^\dagger)|\Omega \rangle \;,
\end{equation}
where  $j_5(x)$ is the divergence of the axial-vector current
\begin{equation}
j_5(x)=(m_b + m_q) : \overline{q}(x)i\gamma_5 b(x): \;,
\end{equation}
and $m_b$ and $m_q$ are the masses of the bottom quark $b$ and the lighter quark $q$, respectively.  We then analytically continue $\Pi(q^2)$ over the complex squared energy s-plane and invoke Cauchy's theorem
\begin{equation}
\frac{1}{2\pi i} \oint_\Gamma P(s)\Pi(s) ds = 0 \;,
\end{equation}
valid for all holomorphic functions $P(s)$, and all closed curves $\Gamma$ which do not encircle a singularity of $\Pi(s)$.  The correlator has singularities only on the positive real axis for $|s| > s_{\textrm{thr}}$, the physical threshold.  Choosing $\Gamma$ to correspond to a circle of radius $|s| = s_0$, along both sides of the cut on the real axis, and using the Schwarz reflection principle, one finds
\begin{equation}
\frac{1}{\pi}\int^{s_0}_{s_\textrm{thr}} {\mbox{Im}} \,[ P(s)\,\Pi(s)] ds = -\frac{1}{2\pi i}\oint_{|s|=s_0}P(s)\Pi(s) ds \;.
\label{eq:SR}
\end{equation}
This leads to a relation between QCD parameters and experimental observables, after invoking quark-hadron duality, i.e. assuming that $\Pi(s)$ in the contour integral is given by QCD if $s_0$ is large enough, i.e.
\begin{equation}
\Pi(s)|_{|s|=s_0}
=\Pi^{\textrm{pQCD}}(s)+\Pi^{\textrm{npQCD}}(s)\;,
\label{eq:PhQCD}
\end{equation}
where we have explicitly separated the perturbative, $\Pi^{\textrm{pQCD}}(s)$, and the non-perturbative, $\Pi^{\textrm{npQCD}}(s)$ QCD contributions. In this case one obtains the FESR
\begin{equation}
\frac{1}{\pi}\int^{s_0}_{s_\textrm{thr}} {\mbox{Im}} \,[ P(s)\Pi^{\textrm{HAD}}(s)] ds = -\frac{1}{2\pi i}\oint_{|s|=s_0} P(s) \Pi^{\textrm{QCD}}(s) ds\;.
\label{eq:SR2}
\end{equation}

The use of non-trivial integration kernels $P(s)$ in FESR was pioneered in \cite{DUAL} in order to account for potential quark-hadron duality violations, e.g. in the Weinberg sum rules. Since then they have been frequently and successfully used  in a variety of QCDSR applications. In particular, Legendre polynomial kernels subject to global constraints have been employed in extractions of 
the chiral condensates from $\tau$-decay data \cite{CHC}, and on the chiral corrections to the Gell-Mann-Oakes-Renner relations \cite{GMOR}. The current most precise determinations of the charm- and bottom-quark masses \cite{charmq}-\cite{bottomq} are also based on FESR involving these kernels. The general purpose of these kernels is to tune the FESR so as to emphasize or quench energy regions where the information is well or poorly known, respectively. In this way systematic uncertainties can be considerably reduced.
\section{Phenomenological Contribution and $P(s)$}
As usual we parametrize the phenomenological correlator with a single pole for the $B_{q}$ meson. For the unknown hadronic continuum we define the physical threshold
\begin{equation}
s_{\text{phys}}=(M_{B^{\ast}}+M_{P_{q}})^{2} \;,
\end{equation}
where $B^{\ast}$ is a vector meson and $P_{q}$
is the lightest pseudoscalar meson with $q=u,d,s,c$ quantum numbers, namely,
$P_{q}=\pi,$ $K,$ $D,$ respectively.  The spectral density can then be written as
\begin{eqnarray}
\rho(s)  & \equiv& \frac{1}{\pi} \; \operatorname{Im}\Pi^{\textrm{HAD}}(s)
=\frac{1}{\pi
}\operatorname{Im}\Pi^{\text{pole}}(s) \nonumber\\[.2cm]
&+&\frac{1}{\pi}\operatorname{Im}%
\Pi^{\text{cont}}(s)\; \theta(s-s_{\text{phys}}) \nonumber\\ [.2cm]
 &=& M_{B_{q}}^{4} \;f_{B_{q}}^{2} \;\delta(s-M_{B_{q}}^{2}) \nonumber\\ [.2cm]
 &+&\rho^{h}
(s)\; \theta\left[(s-(M_{B^{\ast}} + M_{P_{q}})^{2}\right] \;,
\end{eqnarray}
where $\rho^{h}(s)$ is given by the sum over all hadronic intermediate
states with the quantum numbers of $B_{q}$. To minimize the contribution of the unknown hadronic continuum we shall make a judicious choice of $P(s)$.  We choose a   polynomial
\begin{equation}
P_n(s)=a_0+a_1s+a_2s^2+\ldots+ a_ns^n\;,
\end{equation}
and determine the coefficients, $a_i$, subject to the global constraint 
\begin{equation}
\int_{s_\textrm{cont}}^{s_0} s^k P_n(s) ds = 0 \hspace{0.5cm} \forall k\in\{0,\ldots ,n-1\} \;.
\end{equation}
To fix the last coefficient we use an arbitrary overall normalization condition.  The functions $P_n(s)$ are then Legendre polynomials defined in the interval $\left[  s_{\textrm{cont}},s_{0}\right]$,
where we fix $s_{\textrm{cont}}\thicksim s_{\textrm{phys}}$ by demanding maximum duality in the
sense explained below.
The  introduction of this polynomial kernel in the sum rule minimizes the
continuum contribution to the phenomenological side. In fact, to the
extent that $\mathrm{Im}\,\,\Pi ^{\mathrm{\textrm{cont}}}(s)$ can be approximated by an 
$(n-1)$-th degree polynomial, these conditions lead to an exact
cancellation of the continuum contribution on the left hand side of Eq.
(\ref{eq:SR2}). At the same time, the role of the $B_{q}$ pole will be enhanced.
Increasing $n$ increases the OPE truncation error, but this is compensated by increasing the integration radius $|s_0|$. It will turn out that the latter is of reasonable magnitude on the relevant scale, i.e., $m_B^2$.  If, on the other hand, the radius of integration is chosen too
high, the polynomial fit will deteriorate. It is hoped, and actually
confirmed, that there is a wide intermediate region of stability. To find this
region of stability we begin by choosing a value of $n=3,4,5,..$ and then
increasing $s_{0}$ until stability is reached in each case.
In order to make the duality region more pronounced we allow the value of the
threshold $s_{\textrm{cont}}$ to vary around the physical threshold $s_{\textrm{phys}}$, as in \cite{Melikhov:2012db}, and in such a way that the dependence of $f_{B_{q}}$ on $s_{0}$ shows maximum
stability. This means  requiring that the first and second derivatives vanish at the same point. Reassuringly, we find that the value of $s_{\textrm{cont}}$ which maximizes this stability is very close to $s_{\textrm{phys}}$, thus validating this procedure. The contribution of the threshold region is then estimated
separately. We have reasons to believe that if the resulting $f_{B_{q}}$ is
independent of $n$ and $s_{0}$ in some finite region, then the unknown
theoretical and hadronic contributions effectively cancel there. This
follows from general properties of Legendre polynomials \cite{Sansone}, the details of which will appear in a separate work. We are then left with the key relation
\begin{equation}
M_{B_q}^4f_{B_{q}}^2 P_n(M_{B_q}^2) =  -\frac{1}{2\pi i}\oint_{|s|=s_0}  P_n(s)\Pi^{\textrm{QCD}}(s) ds \;.
\end{equation}
\section{Perturbative Contribution}
We now turn to the perturbative part of the QCD correlator, $\Pi^\textrm{pQCD}(s)$.  To calculate $f_{B_s}$ and $f_{B_c}$ we  need to take into account the effect of the lighter quark.  We separate the correlator into its expression in the $m_q = 0$ limit, and  add to it the lighter quark mass corrections
\begin{equation}
\Pi^\textrm{pQCD}(s)=\Pi^\textrm{pQCD}_{m_q=0}(s)+\Pi^\textrm{pQCD}_{m_q}(s)
\end{equation}
For $\Pi^\textrm{pQCD}_{m_q=0}(s)$ we use the high-energy expansion given in \cite{Chetyrkin:2001je}. This is an  expansion to order $O(\alpha_s^2)$
\begin{eqnarray}
\Pi^\textrm{pQCD}_{m_q=0}(s)&=&\Pi^{(0)}_{m_q=0}(s)+\Pi^{(1)}_{m_q=0}(s)\left(\frac{\alpha_s(\mu)}{\pi}\right)\nonumber\\[.2cm]
&+&\Pi^{(2)}_{m_q=0}(s)\left(\frac{\alpha_s(\mu)}{\pi}\right)^2
\end{eqnarray}
and is given to $O(z^{-7})$ in $z=s/\widehat{m}_b^2$, where $\widehat{m}_b$ is the pole mass of the $b$ quark.  For example, to leading order in $\alpha_s$ \cite{Chetyrkin:2001je}
\begin{eqnarray}
\Pi^{(0)\textrm{pQCD}}_{m_q=0}(s)&=& \frac{3}{16\pi^2}(\widehat{m}_b+\widehat{m}_q)^2 s \Bigg\{3+4L_z-8L_z\frac{1}{z}\nonumber\\[.2cm]
&+&(-3+4L_z)\frac{1}{z^2}  
+ \frac{2}{3}\frac{1}{z^3}+\frac{1}{6}\frac{1}{z^4}+\frac{1}{15}\frac{1}{z^5}\nonumber\\[.2cm]
&+&\frac{1}{30}\frac{1}{z^6}+\frac{2}{105}\frac{1}{z^7}\Bigg\}
\end{eqnarray}
where $L_z=-(\ln(-z))/2$ and $\widehat{m}_q$ is the pole mass of the light quark (note that we keep the light quark mass in the pre-factor arising from the divergence of the axial-vector current).  It is well known that the convergence of the QCD correlator is improved if it is expressed in terms of the running quark masses \cite{Jamin:2001fw, Bordes:2004vu}.  We thus write the pole masses in terms of the running masses, $m_b(\mu)$ and $m_q(\mu)$, expand the final expression as a series in $\alpha_s(\mu)$, and truncate at $O(\alpha_s^2)$. As the perturbative expansion is not known to all orders in $\alpha_s(\mu)$ the final result will have some dependence on the renormalization scale, $\mu$.  To improve the convergence we take $\mu=m_b$, which re-sums the $\ln(\frac{m_b^2}{\mu^2})$ terms appearing in the running quark  mass \cite{Bordes:2004vu}.  We later investigate the effect of varying $\mu$, retaining the $\ln(\frac{m_b^2}{\mu^2})$ terms.  The $O(z^{-7})$ truncation has a very good convergence over the range of $s$ used here,  and it introduces a negligible error.\\  

To calculate at $\mu=m_b$, and to investigate the  $\mu$-dependence, we use the  running coupling $\alpha_s(\mu)$ to four-loop order \cite{Bethke:2009jm, vanRitbergen:1997va}, and  running quark masses to four-loops \cite{Vermaseren:1997fq}.
We performed the contour integral in two complementary ways, i.e. analytically and  numerically.  Analytically, we followed the method outlined in \cite{Bordes:2004vu}.  Numerically, we used standard computational methods to perform the contour integral.  The final results  agree to a few MeV, and the error estimates are entirely compatible.
\section{Light Quark Mass Corrections}
The light quark mass corrections were originally obtained to  $O(\alpha_s)$ in \cite{Broadhurst:1981jk, Generalis:1990id}.  These corrections to the full correlator are somewhat unwieldy, but the imaginary part to $O(m^4)$ is presented in a compact form in \cite{Jamin:2001fw}.  For example, the $O(\alpha_s^0)$ term is 
\begin{eqnarray}
 {\mbox{Im}} \,\Pi^{(0)\textrm{pQCD}}_{m_q}(s)&=&\frac{3}{8\pi^2}(m_b+m_q)^2\Big\{2(1-x)m_b m_q  \nonumber \\ [.2cm] 
 &-& 2 m_q^2 - 2 \;\frac{(1+x)}{(1-x)}\frac{m_bm_q^3}{s}\nonumber \\ [.2cm]
 &+&\frac{(1-2x-x^2)}{(1-x)^2}\frac{m_q^4}{s}\Big\}
\end{eqnarray}
where $x=m_b^2/s$.  The $O(\alpha_s)$ term can be found in the appendix of \cite{Jamin:2001fw}. 
In the case of $f_{B_c}$, where the lighter (charm) quark mass correction is almost 10 \%, the $O(m^4)$ terms contribute less than 1 \%, so we are justified in making this approximation.  Of the 10 \% correction, around 4 \% comes from the $O(\alpha_s)$ term, indicating that the convergence is not ideal.  It is, however, good enough for the present level of precision, and in any case it is somewhat included in the determination of the error from  varying $\mu$ in the range 3 - 6 GeV.  In the case of $f_{B_s}$ the light quark mass correction is around 5 \%.  For $f_B$ we take $m_u=m_d=0$. To evaluate the integral of $P_n(s)\Pi^{\textrm{pQCD}}_{m_q}(s)$ we again use the Cauchy integral theorem to rewrite it as the integral over the imaginary part along the real axis, now starting from $s=(m_b+m_q)^2$, the start of the cut, and use Eq. (\ref{eq:SR}). This is purely a mathematical device, unrelated to the discussion leading to Eq. (\ref{eq:SR2}), so the result is exact and no error is introduced due to the poor convergence of the QCD expansion at low $s$.
\section{Non-perturbative Contribution}
The non-perturbative contributions to the correlator can be parametrized through Wilson's operator product expansion.  In our calculation we include the effects of quark and gluon condensates up to dimension six \cite{Broadhurst:1981jk}-\cite{Generalis:1990id}
\begin{eqnarray}
&&\Pi^{\textrm{npQCD}}(s) = (\widehat{m}_b+\widehat{m}_q)^2 \bigg\{ \frac{\widehat{m}_b\left<\bar{q}q\right>}{s-\widehat{m}_b^2}\left(1+2\frac{\alpha_s}{\pi}\right) \nonumber \\ [.2cm]
&&-\frac{1}{12}\frac{1}{(s-\widehat{m}_b^2)}\left<\frac{\alpha_s G^2}{\pi}\right> 
- \frac{\widehat{m}_b\left<\bar{q}\sigma G q\right>}{2}\left[\frac{1}{(s-\widehat{m}_b^2)^2} \right. \nonumber \\ [.2cm]
&& \left. +\frac{\widehat{m}_b^2}{(s-\widehat{m}_b^2)^3}\right] 
 -\frac{8\pi\alpha_s\left<\bar{q}q\right>^2}{27}\left[\frac{2}{(s-\widehat{m}_b^2)^ 2} \right.\nonumber \\ [.2cm]
&& \left. +\frac{\widehat{m}_b^2}{(s-\widehat{m}_b^2)^3}-\frac{\widehat{m}_b^4}{(s-\widehat {m}_b^2)^4}\right]\bigg\}
\end{eqnarray}
Note that we include the effect of the light quark mass only in the pre-factor.  The non-perturbative part contributes around 10 \% to $f_B$, and  it is dominated by the lowest dimensional condensate $\langle \bar{q} q \rangle$. Its value is usually given at 2 GeV, and determined e.g. from the Gell-Mann-Oakes-Renner relation, with the result \cite{GMOR} $\langle \bar{q} q \rangle = (- 267 \pm 5 \;{\mbox{MeV}})^3$.  To run this condensate to the renormalization scale $\mu$ we use the fact that $m_q(\mu)\left<\bar{q}q\right>$ is renormalization group invariant.  For the strange quark condensate, and the mixed condensates we use
\begin{equation}
\left<\bar{s}s\right> =R_{sq}  \left<\bar{q}q\right> \;,
\end{equation}
\begin{equation}
\left<\bar{q}\sigma G q\right> = m_0^2 \left<\bar{q}q\right>\;,
\end{equation}
\begin{equation}
\left<\bar{s}\sigma G s\right> = m_0^2\; R_{sq}\; \left<\bar{q}q\right>\;,
\end{equation}
where  $R_{sq}= 0.6 \pm 0.1$ is from \cite{RSQ}, and $m_0^2 = 0.8 \pm 0.2 \;{\mbox{GeV}}^2$ from \cite{Ovchinnikov:1988gk}. The gluon condensate has been determined from data on the hadronic decays of the $\tau$-lepton \cite{G2}, i.e. 
$\left<\frac{\alpha_s G^2}{\pi}\right> = 0.07 \pm 0.02 \; {\mbox{GeV}}^4.$
  The terms involving  $\left<\frac{\alpha_s G^2}{\pi}\right>$ and $\left<\bar{q}\sigma G q\right>$  both contribute around 0.1 \%, while the term involving  $\langle \bar{q} q \rangle^2$ contributes around 0.001 \%, so we are justified in ignoring higher dimensional operators.  In $f_{B_s}$  the non-perturbative part plays a smaller role, while for $f_{B_c}$ it is negligible.  The $O(\alpha_s)$ correction to the $\langle \bar{q} q \rangle$ condensate term contributes around 1 \% to $f_B$, so this approximation is accurate enough.  
As with the perturbative part, we rewrite the pole mass in terms of the running mass \cite{Gray:1990yh}, expand in powers of $\alpha_s$ and truncate the series at $O(\alpha_s^2)$.  The resulting integral is easily evaluated using Cauchy's residue theorem.
\begin{table}
\begin{center}
\begin{tabular}{c|ccc}
\hline
$n$ & $f_B$ (MeV)& $f_{B_s}$ (MeV)& $f_{B_c}$ (MeV)\\
\hline
3 & 188.7 & 222.2 & 521.9  \\
4 & 186.9 & 221.7 & 525.1 \\
5 & 186.4 & 221.8 & 526.7 \\
6 & 186.2 & 221.9 & 527.6 \\
\hline
\end{tabular}
\caption{Results for the decay constants for different values of $n$.}
\label{tab:DR}
\end{center}
\end{table}
\begin{table}
\begin{tabular}{ccccc}
\multicolumn{5}{r}{Uncertainties (MeV)} \\
\cline{3-5}
\noalign{\smallskip}
 INPUT & VALUE & $\Delta f_B$ &  $\Delta f_{B_s}$  & $\Delta f_{B_c}$                \\
\hline
\noalign{\smallskip}
$O(\alpha_s^2)$ \quad & doubling/removing \quad & $\mp 7$ \quad &\quad $\mp 4$ \quad & \quad $\pm 13$\\
\\
$\alpha_s(M_Z)$ \quad & $0.1184 \pm 0.0007$ \quad &  $\mp 0.5$ \quad & \quad $\mp 0.3$ \quad & \quad $\pm 0.6$\\
\\
$\overline{m}_b(m_b)$ \quad & $4.18\pm 0.03 \;{\mbox{GeV}}$ \quad & $\mp 10$ \quad & \quad $\mp 10$ \quad & \quad $\mp 9$\\
\\
$\overline{m}_s(2\; {\mbox{GeV}})$ \quad & $94.0 \pm 9.0 \;{\mbox{MeV}}$ \quad & \o \quad & \quad $\pm 1.5$ \quad & \quad \o\\
\\
$\overline{m}_c(m_c)$ \quad & $1.278 \pm 0.009 \;{\mbox{GeV}}$ \quad & \o \quad & \quad \o \quad & \quad $\pm 0.6$\\
\\
$ - \langle \bar{q} q \rangle (2 \; {\mbox{GeV}}) $ \quad & $(267 \pm 5)^3\; \mbox{MeV}^3$ \quad & $\mp 2.1$ \quad & $\mp 1.1$  \quad & \quad  - \\ 
\\
$\langle \alpha_s G^2/\pi \rangle$ \quad & $0.07 \pm 0.02 {\mbox{GeV}}^4$ \quad & $\pm 0.5$ \quad & $\pm 0.3$ \quad &  \quad $\pm 0.2$\\
\\
$R_{sq}$ \quad & $0.6 \pm 0.2$ \quad & \o \quad & $\pm 1.7$ \quad & \quad \o \\
\\
$m_0^2$ \quad & $0.6 \pm 0.2$ \quad & $\mp 0.1$ \quad & - \quad & \quad - \\
\\
$M_B$ \quad & $5279.58 \pm 0.17 \; {\mbox{MeV}}$ \quad & - \quad & \o \quad & \quad \o\\
\\
$M_{B_s}$ \quad & $5366.77 \pm 0.24 \; {\mbox{MeV}}$ \quad & \o \quad & -\quad & \quad \o\\
\\
$M_{B_c}$ \quad & $6277 \pm 6 \; {\mbox{MeV}}$ \quad & \o \quad & \o \quad & \quad $\pm 1.7$\\
\end{tabular}
\caption{\scriptsize{Input values of parameters (see text for references) and their contribution to the  uncertainties in the decay constants. A  \o $ $  indicates not applicable.}}
\label{tab:inp}
\end{table}
\section{Results}
\begin{figure}
 \begin{center}
 \includegraphics[height=2.7in, width=4.4in]{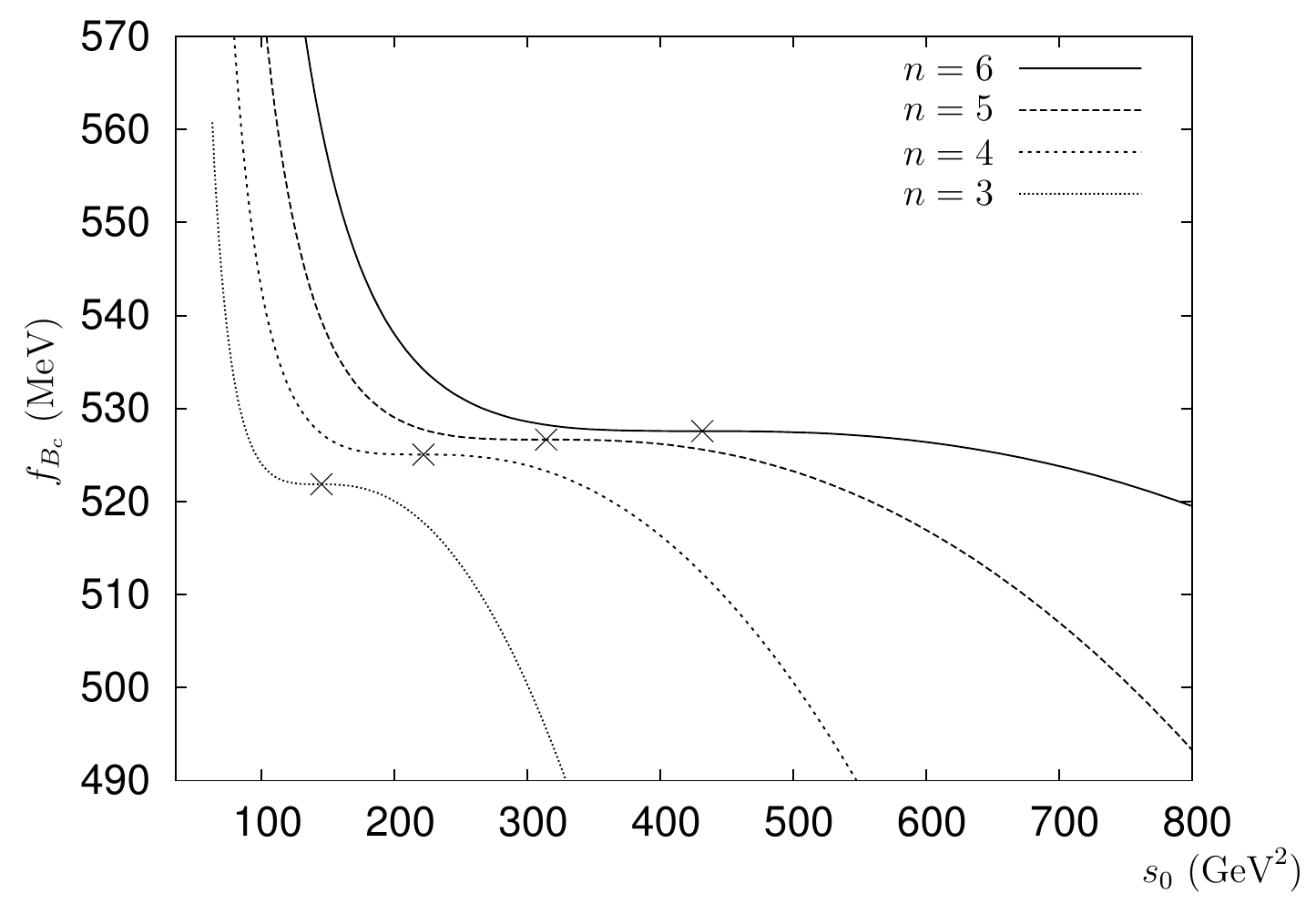}
 \caption{The decay constant $f_{B_c}$ as a function of the integration contour radius $s_0$, for different values of $n$.  The crosses denote the inflection point where the prediction is extracted.}
 \label{fig:fBc}
 \end{center}
 \end{figure}

\begin{figure}
 \begin{center}
 \includegraphics[height=2.7in, width=4.4in]{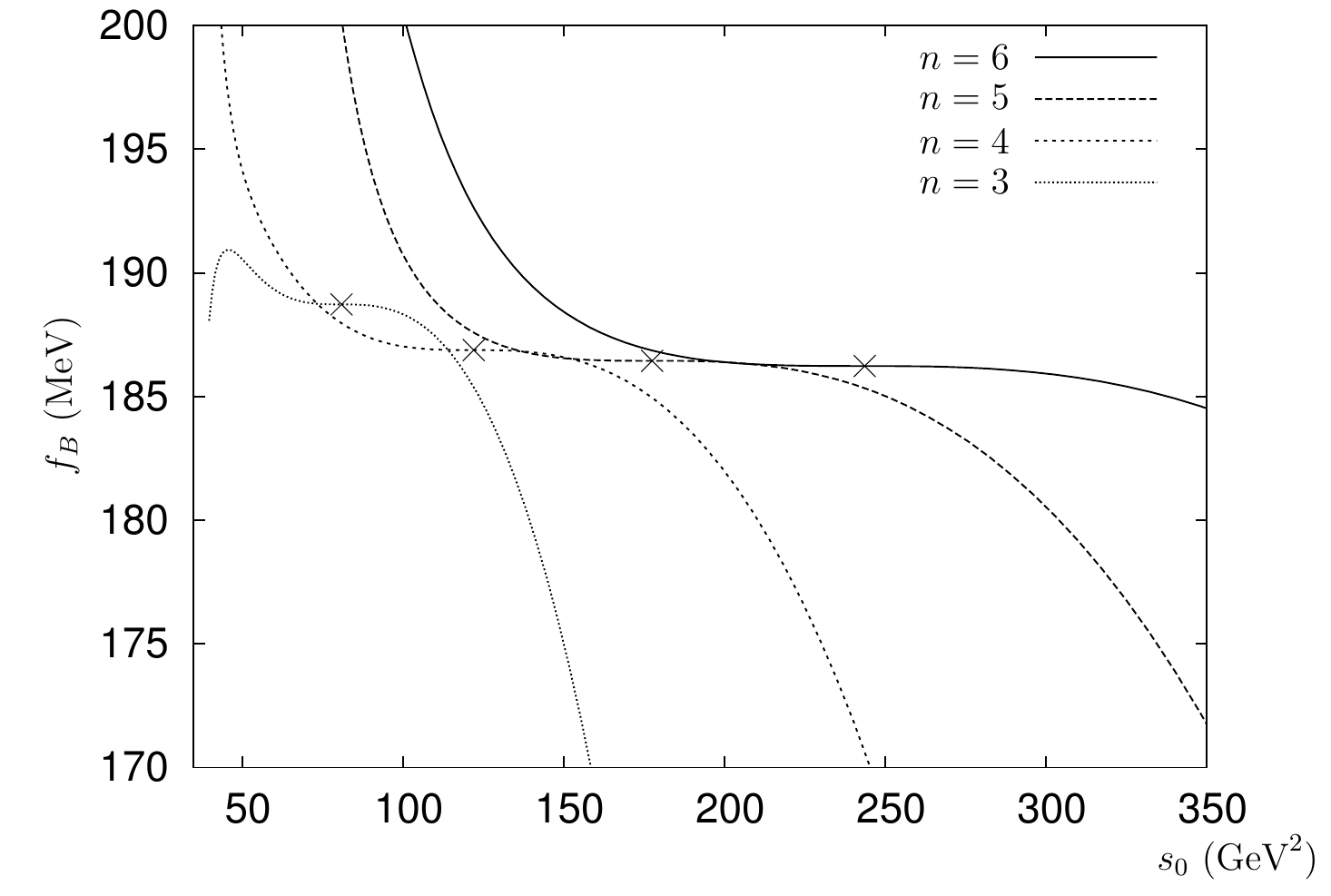}
 \caption{The decay constant $f_{B}$ as a function of the integration contour radius $s_0$, for different values of $n$.  The crosses denote the inflection point where the prediction is extracted}
 \label{fig:fB}
 \end{center}
 \end{figure}

\begin{figure}
\begin{center}
\includegraphics[height=3.0in, width=4.5in]{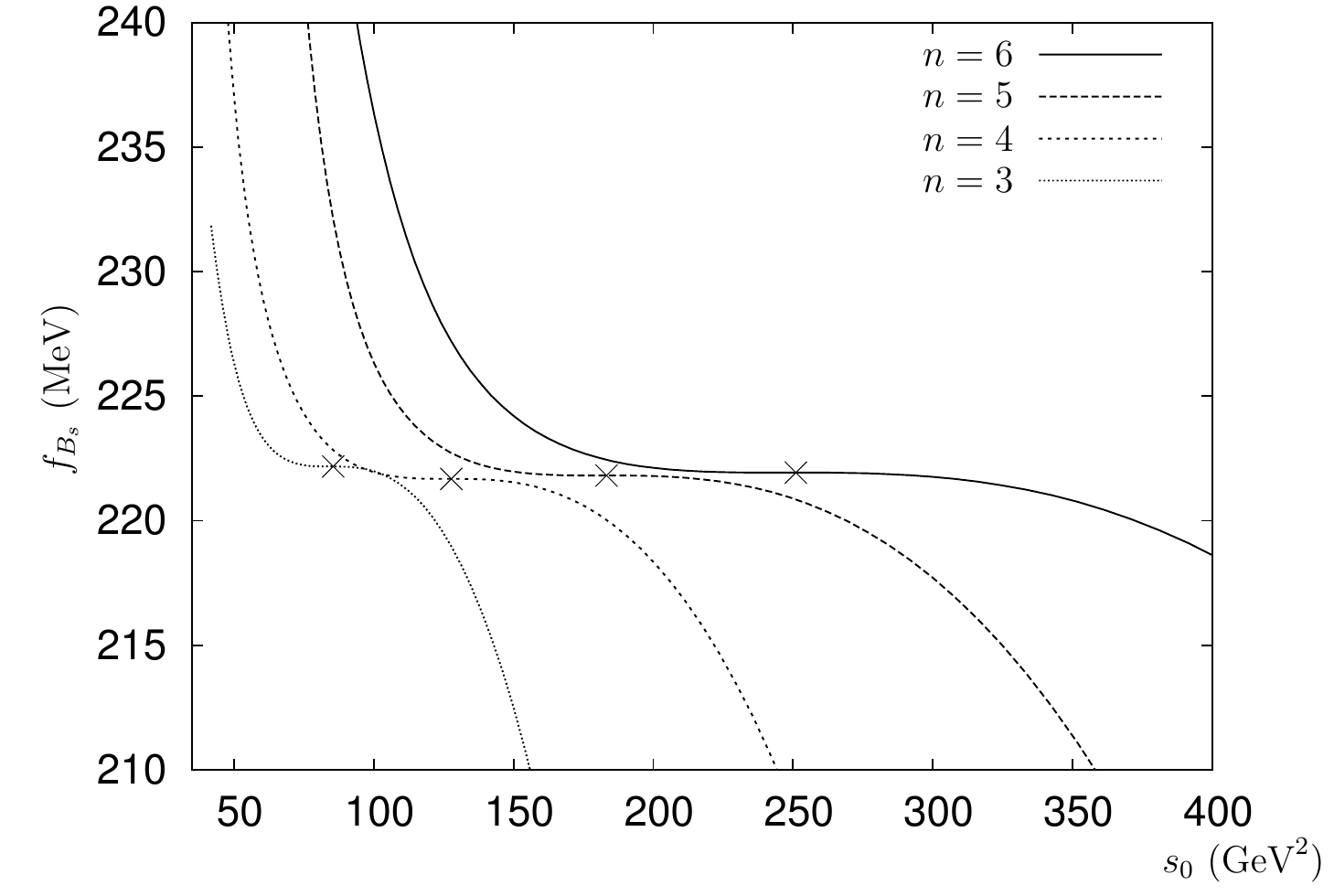}
\caption{The decay constant $f_{B_s}$ as a function of the integration contour radius $s_0$, for different values of $n$.  The crosses denote the inflection point where the prediction is extracted}
\label{fig:fBs}
\end{center}
\end{figure}

Given the complete QCD expansion, and the correct hadronic continuum, the final results would be independent of the two parameters $n$ and $s_0$. In practice, though, there is a finite region in this parameter space where results are stable. The wider this region, the more accurate the results.
 As explained earlier, we allow $s_\textrm{cont}$ to vary so that the first and second derivatives of $f_{B_q}$ with respect to $s_0$ vanish at some point.  We then extract our prediction for each $n$ from this inflection  point.  In figures \ref{fig:fBc} - \ref{fig:fBs} we show $f_{B_c}$, $f_{B}$ and $f_{B_s}$ for different choices of $n$.  It can be seen from these figures that for each $n$ there is range of $s_0$ where the decay constant is virtually independent of $s_0$.  The extension of this range increases with increasing $n$, as the unknown contributions are better modeled by higher order polynomials. These plateau agree very well for different $n$ and exhibit very good converge with increasing $n$, as shown in table I, and in the figures.  The excellent stability in $s_0$, combined with the good convergence in $n$, indicates that the contributions from the unknown hadronic continuum, and the unknown part of the QCD correlator should be small.  We conservatively estimate systematic errors as the difference between the $n=3$ and $n=6$ result, giving $3$ MeV, $1$ MeV and $6$ MeV for $f_{B}$, $f_{B_s}$ and $f_{B_c}$, respectively. 
From the  stability criterion we find e.g. with $n=6$, and for $B,$ $B_{s}$ and $B_{c}$, respectively, the values  $s_{\textrm{cont}}= 31.3 \;\textrm{GeV}^{2},\, 33.0 \; \textrm{GeV}^{2}, \,50.6 \;\textrm{GeV}^{2}$. These
are very close to the values of the actual physical thresholds $ s_{phys} = 29.9$ $\textrm{GeV}^{2}$, $33.8$ $\textrm{GeV}^{2}$,
 $51.7$ $\textrm{GeV}^{2}$.
To understand the effect of the variation of $s_{\textrm{cont}}$ we estimated the
contribution of the continuum near the physical threshold in the determination
of $f_{B}$, where some phenomenological information is available. We considered
the phase space contribution of the $B\pi\pi$ and~$B^{\ast}\pi$ intermediate
states near threshold, with form factors taken from chiral perturbation theory
\cite{(CAD)}-\cite{BODms} and LQCD \cite{(BenS)},
respectively. In both cases the contributions are negligible compared with the
dominant pole contribution of the $B$ meson. We also investigated the
contribution of a possible  excited $B$ state such as the $B_{1}(5721)$
predicted in \cite{(COL)} from heavy quark effective theories. The
contribution of such a resonance will increase  $f_{B}$  by $4-5$~MeV, which must be considered as part of the systematic
error. We expect a similar situation for $f_{B_{s}}$ and $f_{B_{c}}$.
The final result should be independent of the renormalization scale, $\mu$.  However, due to the various truncations of expansions in $\alpha_s(\mu)$   some $\mu$ dependence is introduced, which is roughly related to the convergence of the asymptotic expansion.  To estimate the error we either remove or double the $O(\alpha_s^2)$ correction.  This is shown in the first line of table II.  As an alternative one could vary $\mu$ in the range 3 - 6 GeV, which changes $f_{B_c}$, $f_{B}$ and $f_{B_s}$ by up to 10 MeV, 8 MeV, and 10 MeV, respectively.  As these two errors are highly correlated we use only the error estimated from the $O(\alpha_s^2)$ correction.\\
Of the various inputs, the largest error comes from the running mass of the bottom quark, $\overline{m}_b(\overline{MS})$.  We use the Particle Data Group world average  $\overline{m}_b(m_b) = 4.18 \pm 0.03 \; {\mbox{GeV}}$ \cite{Beringer:1900zz}, very close to the
 most recent and accurate value
 $\overline{m}_b(m_b) = 4.171 \pm 0.009 \; {\mbox{GeV}}$ \cite{bottomq}, \cite{REVIEWmq}.  
 In table II we show the change in the decay constants when $m_b(\overline{MS})$ is changed by $\pm \, 1 \sigma$.  Despite the relatively large uncertainty in the $s$-quark mass \cite{BODms}, \cite{REVIEWmq}, $\overline{m}_s (2\; {\mbox{GeV}}) = 94 \pm 9 \;{\mbox{MeV}}$, it only has a small effect on $f_{B_s}$, reflecting the relative smallness of this correction.  The uncertainty in the larger $c$-quark mass $\bar{m}_c(m_c) = 1.278 \; \pm 0.009\; {\mbox {GeV}}$ \cite{REVIEWmq}, \cite{BODmc} has also a small impact on $f_{B_c}$. 

Although the non-perturbative part contributes only around 10 \% to $f_{B}$, and less to $f_{B_s}$, the errors of the condensates are not all negligible.  The uncertainty in the quark condensate, $\langle \bar{q}q\rangle$(2 GeV), affects $f_{B}$ and $f_{B_s}$ by around 2 MeV and 1 MeV, respectively (the effect on $f_{B_c}$ is negligible).  The relatively large uncertainty in $\langle \alpha_s G^2/\pi \rangle$ has a small effect, while the errors in the higher dimensional condensates have a negligible effect, reflecting the fact that the quark condensate dominates the non-perturbative contribution.  The uncertainty in $m_0^2$ has a non-negligible effect only for $f_B$.  There is some disagreement in the literature on the value of $R_{sq}$. A recent LQCD result \cite{McNeile:2012xh} claims   $R_{sq} = 1.08 \pm 0.16$, which is in conflict with almost all determinations leading to $R_{sq} < 1$ (for an exception see \cite{MALT}). This value
 translates into $f_{B_s} = 230$ MeV. The uncertainty in the strong coupling constant, $\alpha_s(M_Z)$, has a small effect on the decay constants.  The masses of the lowest lying pseudoscalar resonances are now very well known \cite{Beringer:1900zz}, and the only non-negligible error arises from the uncertainty in $M_{B_c}$. Adding these errors in quadrature we find
\begin{equation}
f_{B_c} = 528 \pm 9 \pm 17 \textrm{ MeV} \;,
\end{equation}
\begin{equation}
f_B = 186 \pm 11 \pm 9 \textrm{ MeV} \;,
\end{equation}
\begin{equation}
f_{B_s} = 222 \pm 11 \pm 4 \textrm{ MeV} \;,
\end{equation}
where the first error comes from the inputs and the second is systematic, arising from the dependence on $n$, the $O(\alpha_s^2)$ truncation, and the light quark mass expansion.
The ratio of decay constants $\frac{f_{Bs}}{f_B}$ (which is unity in the chiral limit) can be determined quite accurately since many of the errors are correlated.  We find
\begin{equation}
\frac{f_{B_s}}{f_B} = 1.19 \pm 0.02 \pm 0.04 \;,
\end{equation}
where again the first error comes from the inputs and the second is systematic.
The effect of including the light quark mass terms in the perturbative expansion, $\Pi^{(0)\textrm{pert}}_{m_{q}} (s)$, is similar in size to the $O(\alpha_s^2)$ correction for $f_{B_s}$ but is considerably larger for $f_{B_c}$.  The mass corrections increase $f_{B_s}$ by $11\; {\mbox{MeV}}$, whereas $f_{B_c}$ is increased by 45 MeV.
\section{Conclusion}
In this paper we discussed a new QCDSR determination of the heavy meson decay constant $f_{B_c}$, and an update of $f_B$ and $f_{B_s}$. We used Legendre polynomial integration kernels, and proposed a new technique  allowing the continuum threshold, $s_\textrm{cont}$, to vary. This variation leads to an effective  suppression of the unknown hadronic continuum, and  the unknown terms in the QCD correlator. For the latter we used an $O(\alpha_s^2)$,  and $O((\widehat{m}_b^2/s)^7)$ expansion of the pseudoscalar two-point function, supplemented by $O(\alpha_s)$, and $O(m_q^4)$ lighter quark mass corrections. Non-perturbative terms were incorporated through the OPE  up to dimension six. For all three decay constants we found excellent stability over a wide range of $s_0$ and good convergence in $n$, the order of the Legendre polynomials.\\
 
The result for $f_{B_c}$ agrees with some LQCD, and  a few  determinations in other frameworks reviewed in \cite{Jones}, but it is  around 20 \% higher than that obtained in \cite{McN12a} from LQCD. However, it has been argued \cite{Benbrik} that LQCD results for this constant may underestimate its value by some 23\%. It must be pointed out that results for $f_{B_q}$ are very sensitive to the mass of the $q$-quark. For instance, in the case of $f_{B_s}$ the impact of a strange-quark mass $m_s \simeq 100\; {\mbox{MeV}}$ results in a 20\% increase in $f_{B_s}$ relative to $f_B$. For $f_{B_c}$, with $m_c \simeq 1.3\; {\mbox{GeV}}$,  this rough argument would imply an increase over $f_B$ of some 200 \%.
 New and more accurate determinations should resolve this issue. It should be stressed in closing that for $f_B$ and $f_{B_s}$ our results are perfectly consistent within errors with LQCD and recent experimental results.
\section{Acknowledgements}
This work was supported in part by the Alexander von Humboldt Foundation (Germany), NRF (South Africa), an MECD Salvador de Madariaga
Program grant, by MEC and FEDER (EC) under grant FPA2011-23596 and GV under grant
PROMETEO2010-056.

\end{document}